\documentclass[11pt]{article}

\usepackage[a4paper,margin=1in]{geometry}
\usepackage{float}
\usepackage{pifont}
\usepackage{mathtools}
\usepackage{hhline}
\usepackage{booktabs}
\usepackage{array}
\usepackage{siunitx}
\usepackage{xcolor}
\usepackage{listings}
\usepackage{tabularx}
\usepackage{placeins}
\usepackage{caption}
\usepackage{CJKutf8}
\usepackage[authoryear]{natbib}
\setcitestyle{aysep={}}
\let\cite\citep
\usepackage[hidelinks]{hyperref}

\title{A Large Language Model-Based Framework for Automated Game Quest Testing}
\author{%
Enhong Mu$^{1}$ \quad Nianyu Li$^{3}$ \quad Jinyu Cai$^{2}$\\
Mingyue Zhang$^{1,*}$ \quad Kenji Tei$^{4}$ \quad Jialong Li$^{2}$\\[0.8em]
\small $^{1}$Southwest University, Chongqing, China\\
\small $^{2}$Waseda University, Tokyo, Japan\\
\small $^{3}$Zhongguancun Lab, Beijing, China\\
\small $^{4}$Institute of Science Tokyo, Tokyo, Japan\\[0.5em]
\small \texttt{muenhong@email.swu.edu.cn}; \texttt{li\_nianyu@pku.edu.cn}; \texttt{bluelink@toki.waseda.jp}\\
\small \texttt{myzhangswu@swu.edu.cn}; \texttt{tei@comp.isct.ac.jp}; \texttt{lijialong@fuji.waseda.jp}\\
\small $^*$Corresponding author: \texttt{myzhangswu@swu.edu.cn}
}
\date{}

\begin{document}

\maketitle

\begin{abstract}
Quests are central to the experience of many role-playing, simulation, and sandbox games. In modern titles, quests are often complex and frequently updated, which substantially increases testing difficulty. Traditional script-based testing is inefficient and struggles to keep pace with rapid iteration, while existing agent-based methods can discover defects but require time-consuming training and generalize poorly to new quests.
To address these challenges, this paper proposes an automated game quest testing framework based on large language models (LLMs), called LLMEGQT. The framework uses the natural language understanding and reasoning abilities of LLMs to parse quest descriptions into sub-goals and transform them into executable actions; combines multi-round generation with Monte Carlo sampling to build a state-action library and a state transition graph; and finally applies path-planning algorithms to generate execution paths. The method enables rapid quest completion verification and improves coverage over a broader range of states and paths. Experimental results show that the framework achieves higher task completion rates and predefined target coverage across multiple complex quest scenarios, improving the automation of game quest testing.
\end{abstract}

\noindent\textbf{Keywords:} Game Quest Testing; Large Language Model; Goal Planning; Automated Testing

\section{Introduction}
{
In recent years, the video game industry has grown steadily and has become an important force in the entertainment sector, with major economic and social impact \cite{antczak2023finance,goh2023unravelling,cohendet2018local}.
According to Newzoo's global game market report, the compound annual growth rate of the global video game market during 2021--2025 will remain at 5.0\%, reaching USD 218.8 billion in 2024 \cite{newzoo2024}. A 2024 PwC report also states that by 2028, revenue from the global video game industry is expected to exceed USD 300 billion and account for 9\% of the overall entertainment industry \cite{pwc2024}.
Across role-playing, simulation, and sandbox games, quests organize player actions into sequences of objectives and state-dependent interactions. Their multi-stage structure and frequent revision make quest logic an important and challenging target for automated testing \cite{dickey2007game,report-game2023,cnga_report_2024}.

Within the quest-driven game development lifecycle, two characteristics differ from traditional software. First, quest design is often expressed as natural-language stories, objectives, and progression rules before being implemented as executable game logic. Low-quality or defective quests can seriously damage player experience and may even cause substantial financial losses for game companies \cite{bbc_news_2024,jie2024alanguage}.
Second, iteration is rapid. Game companies frequently create, maintain, and adjust large numbers of quest versions; large commercial games may process multiple internal builds and hundreds of quest modifications per week \cite{yuechen2020regression,Cristiano2021Survey,Berg_Marklund2019}.

During game quest testing, verifying whether a quest can be completed successfully is a primary goal. This rapid confirmation is similar to smoke testing in traditional software development, whose core purpose is to ensure that key functionality can execute normally \cite{9787844,Herbold2022,DBLP:conf/icst/CannavacciuoloM22,mcconnell1996rapid}. In game development practice, this approach provides a quick check of basic quest executability \cite{jie2024alanguage,Cai2026SmokeTG}. On this basis, broader state and path coverage can expose additional potential defects. Therefore, quest completion and coverage are complementary objectives in game quest testing.

However, achieving this goal in practice is challenging. The most direct approach is to ask human testers to read quest documents, dialogues, and game instructions, arrange action sequences that can complete the quest, and encode them as executable scripts. Because quests can be complex and frequently updated, manual scripting has high labor and maintenance costs.
In addition, some prior work has introduced agent-based techniques into game testing. With the help of domain knowledge, trained agents can complete specific tasks and discover game defects. Some studies train human-like agents to imitate testers or players and thereby generate test sequences \cite{zook2019automatic,ariyurek2019automated,ariyurek2020enhancing}; others train agents to execute specific testing tasks through carefully designed reward mechanisms \cite{loubos_automated_2018,pfau2017automated,20191438}; still others use evolutionary algorithms to optimize agents for finding potential issues \cite{zheng2019wuji,20226387}. However, these approaches have clear limitations: first, agent training or evolution is usually tailored to a specific task, so when the task changes the agent must be retrained or re-evolved; second, training and evolution are usually time-consuming and inefficient. These problems limit their practical application to large-scale game quest testing.

The key to efficient game quest testing lies in automating the conversion from documents to action sequences. In fact, the core work of human testers is precisely to extract and arrange executable action sequences from task-related documents.
Based on this observation, we use large language models (LLMs) to assist this conversion process. Their natural-language understanding and planning capabilities make them a promising basis for parsing quest descriptions, inferring task objectives, and proposing action plans \cite{minaee2024largelanguagemodelssurvey,10.1145/3686803,Cai2026SAGE,Mu2026KGPlaytesting,Mu2026Synergizing}.

To this end, we propose an automated game quest testing framework, Large Language Model-Enhanced Game Quest Testing (LLMEGQT). The framework uses LLMs to parse quest text descriptions and game environment observations, generate action sequences for completing quests, and combine this with an online planning mechanism to optimize action selection according to environment feedback. In this way, it can both quickly verify whether a quest can be completed successfully and improve test coverage through exploration over a broader range of states and paths, thereby revealing potential defects.
Specifically, the automated testing agent first uses an LLM to decompose a quest (for example, making a tomato salad) into several sub-goals (such as obtaining ingredients, slicing, and plating), and then translates these sub-goals into executable in-game action sequences. By generating executable sequences multiple times, the agent obtains a set of candidate sequences. On this basis, repeated Monte Carlo sampling is used to explore the game environment and collect state-action-state tuples recording the current state, executed action, and resulting next state. These tuples form a transition library and an explored-state graph. A $k$-shortest-path algorithm then extracts efficient candidate paths from the initial state to the quest completion state; these paths support quest completion checks and coverage-oriented exploration within the observed graph.
}

The main contributions of this paper are as follows:
\begin{itemize}
    \item We propose an LLM-based automated game quest testing framework that realizes the automatic conversion from quest documents to action sequences, effectively supporting quest completion verification while also expanding test coverage.
    \item We combine multi-round generation with Monte Carlo sampling to construct a transition library and explored-state graph, then use $k$-shortest-path planning to derive efficient candidate test paths.
    \item We validate the method on multiple quest scenarios spanning a linear RPG, a cooking simulation, and an open-world sandbox, demonstrating its applicability to different interactive game environments.
\end{itemize}

The rest of this paper is organized as follows. Section~2 reviews related work on game testing and LLM-based agents. Section~3 formally defines the problem. Section~4 presents the proposed LLMEGQT framework in detail. Section~5 describes the experimental setup and discusses the results. Finally, Section~6 concludes the paper and outlines directions for future work.

\section{Related Work}

\emph{Importance and challenges of game testing.}
Game testing is a critical stage in the lifecycle of game software. Its goal is to ensure that games run stably, provide complete functionality, and deliver a good user experience before release and after subsequent updates \cite{DBLP:journals/sbcjis/DuarteMDNE24}. Compared with ordinary commercial software, game software must also pay special attention to gameplay logic, interactive interfaces, and performance issues \cite{DBLP:conf/dsa/ZhangZAI23}. In quest-driven games, multi-stage task structures and frequent updates further increase testing difficulty.

\emph{Scripted and traditional automated testing methods.}
In relatively simple and fixed game scenarios, preset scripts are often used for automated testing \cite{spronck2006adaptive,yuechen2020regression}. Testers write executable action scripts based on quest documents and possible player behaviors. However, as quest complexity increases, the cost and maintenance overhead of writing scripts rise sharply, making the process inefficient.
In addition, GUI testing is often used as a comparison point for game testing \cite{lan2024deeply,memon2013automated}. Traditional GUI testing mainly verifies the functional integrity and interaction correctness of interface elements, such as button responses, input validation, and page navigation. These tasks are usually deterministic and have limited state spaces, so they can be handled through script recording, replay, and assertion checking. In contrast, game quest testing involves dynamic state transitions and multi-stage quest logic that traditional GUI techniques do not directly address.

\emph{AI-driven testing methods.}
To improve automation, researchers have introduced artificial intelligence methods. One representative category is imitation learning, where agents are trained to imitate human behavior to complete testing tasks \cite{zook2019automatic,ariyurek2019automated,ariyurek2020enhancing}. However, such methods depend heavily on the quality of human data and have limited generalization ability, making them difficult to apply to new scenarios. Another category uses carefully designed reward functions to drive agents to complete testing tasks \cite{loubos_automated_2018,pfau2017automated,20191438}. Yet reward design is usually scenario-specific, and a general reward function is hard to define, which still limits applicability.

\emph{Exploratory testing in complex environments.}
In more complex task environments, researchers have proposed a series of exploratory testing methods. The goal of these methods is to discover latent errors through diverse exploration. A representative example is the Wuji framework \cite{zheng2019wuji}, which uses evolutionary algorithms to generate diverse agents and thereby improves the chance of exploring new states and triggering errors. Other studies use agents with different preferences to explore the environment \cite{guerrero-romero_using_2018}, but these preferences are usually predefined by humans and lack dynamic adaptability.
In addition, graph-based methods \cite{ammanabrolu_playing_2018,debnath2019review} and tree-based methods \cite{auer2002finite,silver_mastering_2017} have also been applied to game exploration to support path planning and state rollout. Since game quests are often expressed in text, some studies have also explored text classification \cite{joulin_bag_2017,aggarwal_survey_2012} and other natural language processing techniques \cite{he_deep_2015,chen_learning_2011} to assist testing.
Although these methods have made progress in expanding test coverage, they remain insufficient in terms of stability and efficiency for verifying quest completability.

\emph{Positioning of this work.}
In summary, existing work has advanced functional validation and exploratory testing, but remains insufficient for combining quest completion checks with coverage-oriented exploration. Successfully executing a path from the initial state to a completion state establishes the executability of that path, while failure or additional path exploration can expose possible problems without proving the correctness of the entire quest. We therefore combine language-guided action generation, state-space modeling, and path planning to support both objectives.

\section{Problem Definition}

{
For game quest testing, the core objective is to quickly verify whether a quest can be completed successfully \cite{9787844,Herbold2022}. In this paper, we achieve automated quest completion verification by constructing an intelligent testing agent. The task can be described as follows: create an agent that can interact with the game environment and complete the quest autonomously, thereby confirming whether the main game quest can be completed as expected during execution. We model the game environment as a system with the Markov property, so the interaction between the agent and the game environment can be formalized as a Markov decision process (MDP). During decision making, the agent relies not only on the environment state but also on the textual quest description. For later discussion, we give its formal definition below.

The interaction model between the agent and the game is defined as:
} $\mathcal{M}=<S, A, T, D, C>$, where $S$ is the set of states; $A$ is the set of actions; $T : S \times A \rightarrow S$ denotes the state transition function, where $T(s_t,a_t)=s_{t+1}$ means that at time step $t$, when the state is $s_t$ and the testing agent takes action $a_t$, the system transitions to state $s_{t+1}$ at the next time step; $D$ is the set of game quest descriptions, and $d_t\in D$ denotes the quest description text produced by the game environment at time step $t$; $C$ denotes the set of quest completion states. {Here, the time-step subscript $t$ is mainly used to express the temporal order of states, actions, and quest descriptions, that is, when $t>t'$, $s_t$/$a_t$/$d_t$ must appear after $s_{t'}$/$a_{t'}$/$d_{t'}$.}
In this paper, we consider a deterministic state transition function, but the framework is also applicable to uncertain transition functions.

In the game environment, at time step $t$ the intelligent testing agent receives the current quest state information $s_t\in S$ and the quest description $d_t\in D$. The state information includes user interface information, environment information (for example, the positions of non-player characters and in-game weather conditions), player status (for example, health and equipment), quest progress, and other game-related data. Based on $s_t$ and $d_t$, the intelligent testing agent selects an appropriate action $a_t \in A$ and executes it in the game environment, and then receives the next state $s_{t+1}$ and quest description $d_{t+1}$. In most cases, the quest description $d_t$ appears at the beginning of the game (that is, when $t=0$), while it may be empty at later time steps. This process repeats until the quest is completed, namely when $s_t\in C$.

The action-selection policy of the intelligent testing agent is defined as $\pi : S \times A \rightarrow [0,1]$, where $\pi(s,a)=p$ means that the probability of taking action $a$ in state $s$ is $p$.
The core objective of testing is to find an optimal policy such that, under this policy, the intelligent testing agent can complete the quest in as few time steps as possible, thereby verifying quest completion. Meanwhile, to improve testing effectiveness, the agent should also explore more states and paths to some extent, so as to expand test coverage and better identify possible defects.

\section{Method}

This paper proposes an LLM-based goal planning and action planning method for automatic quest completion verification and test coverage expansion. As shown in Fig. \ref{fig:enter-label}, the method consists of the following key steps:
\begin{figure}[h!tp]
    \centering
    \includegraphics[width=0.92\linewidth]{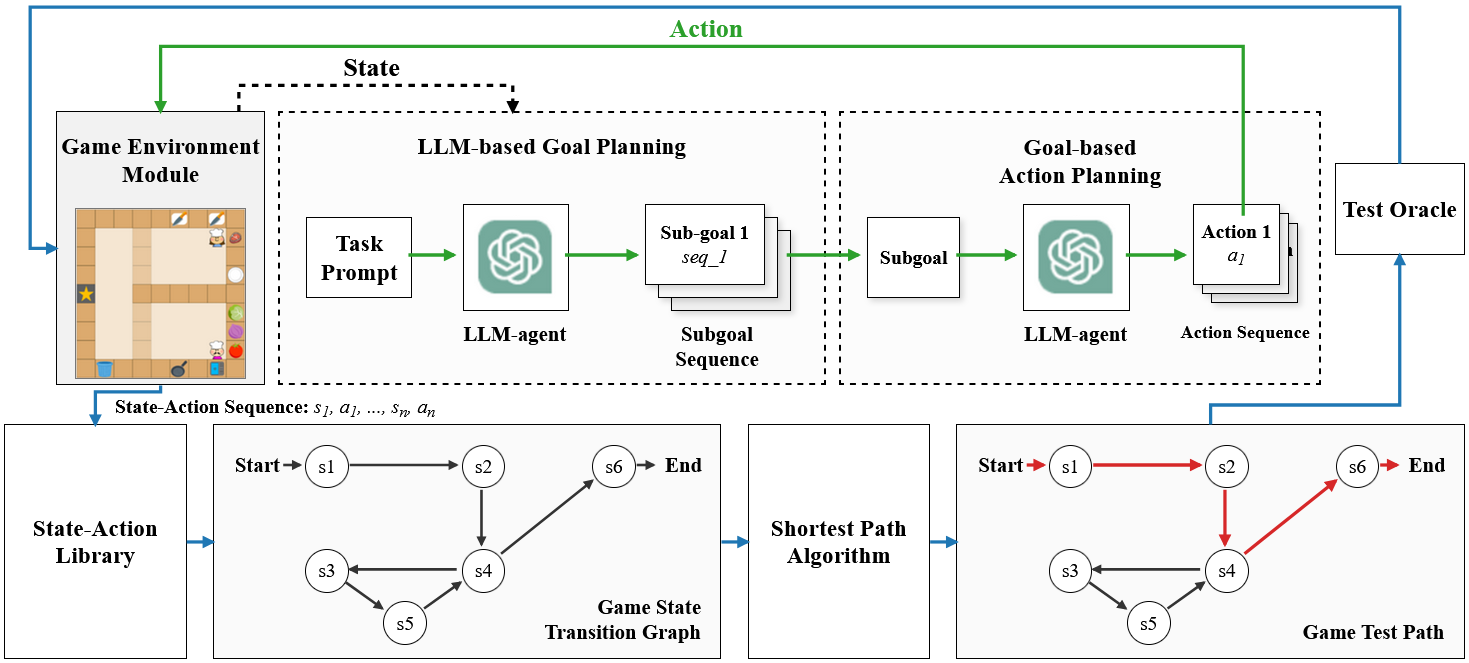}
    \caption{Overview of the LLMEGQT Framework}
    \label{fig:enter-label}
\end{figure}
\begin{itemize}
    \item \textbf{LLM-based goal planning} for quest parsing and key sub-goal generation: the LLM parses the game quest description, extracts key elements, and decomposes the quest into a series of key sub-goals. Through natural language understanding, the LLM generates diverse key sub-goal paths that provide multiple possible guides for quest execution.
    \item \textbf{Goal-based action planning} for action sequence generation: after determining the key sub-goals, the system generates a corresponding action sequence for each sub-goal, including basic operations such as character movement (up, down, left, right), jumping, attacking, and interaction. By combining the LLM with game rules, the system automatically plans the best combination of actions to achieve each sub-goal.
    \item \textbf{Extraction of a state-action library} for game execution and data collection: during action execution, the system records all game states and their corresponding actions to form a state-action library. This library stores state changes and their associated actions and supports the subsequent construction of the state transition graph.
    \item \textbf{Construction of the game state transition graph} for modeling state transitions: based on the state-action library, the system builds a game state transition graph. Nodes represent specific game states, and edges represent actions that cause transitions between states. To expand coverage, the framework combines Monte Carlo sampling with the randomness of LLM generation and executes the above steps multiple times to obtain more samples of the state-action library, thereby improving the graph and more comprehensively modeling the quest state space and transition relations.
    \item \textbf{Candidate-path computation} for path optimization: on the state transition graph, a $k$-shortest-path algorithm derives multiple efficient paths from the initial state to the completion state for subsequent execution and coverage analysis.
\end{itemize}
Through these steps, the proposed method parses game quests, plans concrete completion paths, and uses selected candidate paths to guide efficient execution while covering predefined behavioral targets along explored paths.

\subsection{LLM-Based Goal Planning}
In LLM-based goal planning, we use the natural language understanding and reasoning abilities of the LLM to analyze textual quest descriptions and extract key sub-goals. To improve reasoning performance, the method combines chain-of-thought (CoT) prompting with reflection.

Specifically, we use the state information $s_t$, quest description text $d_t$, and prompt instructions ($\mathcal{I}$) as input to the LLM, and adopt a sequential iterative generation strategy with a total of $n$ rounds. In each round, the LLM first generates an initial sub-goal sequence, then reflects on it and revises it, and finally outputs the revised sequence. All revised sequences from the rounds, $\{seq_1, \dots, seq_{i-1}\}$, form a candidate set that provides diverse task planning options. To ensure diversity, the prompt explicitly constrains the current sequence $seq_i$ not to duplicate any previously generated sequence. The parameter $n$ is the upper bound on the number of iterations and can be set according to task complexity and computational cost.

Within a single sub-goal sequence, each element is represented by the following triplet:
\begin{itemize}
    \item \textbf{Key sub-goal} (\(g_t\)): at each step, the LLM generates the $t$-th sub-goal based on the current state \(s_t\) and quest description text $d_t$. This sub-goal indicates the direction of the current stage of the task.
    \item \textbf{State reasoning} (\(s'_t\)): based on the current state \(s_t\) and sub-goal \(g_t\), the LLM reasons about the expected state \(s'_t\) after executing the sub-goal, using CoT to improve reasoning accuracy.
    \item \textbf{Subsequent guidance} (\(p_t\)): based on the inferred state \(s'_t\) and the quest description \(d_t\), the LLM generates guidance \(p_t\) for the subsequent task, which is used to determine the next sub-goal.
\end{itemize}

This structured generate-reflect-revise process consistently improves the planning capability and execution efficiency of LLMs on complex tasks, and increases the accuracy and reliability of reasoning. Ultimately, the goal planning method outputs a sub-goal sequence $<g_t,g_{t+1},...,g_{t+M}>$ that describes the key steps required to complete the quest, where $g_{t+M}$ denotes the final completion goal. For example, in the kitchen game, sub-goals may include ``go to $x$ to fetch an item,'' ``slice an item with another item,'' and ``produce an item using another item.''

\subsection{Goal-Based Action Planning}
We design a goal-based action planning method whose input is the sub-goal sequence generated by the LLM, $<g_t, g_{t+1}, ..., g_{t+M}>$, and whose output is a concrete action sequence that realizes these sub-goals. For example, for the sub-goal ``go to $x$ and take an item,'' the corresponding actions may be ``move up,'' ``move down,'' and ``pick up the item.'' Specifically, after receiving the sub-goal sequence generated by the LLM, the system further decomposes each sub-goal into executable actions. These actions are basic operations that players can perform in the game, such as movement, jumping, attacking, and interaction. For the sub-goal ``go to $x$ and take an item,'' the system must plan a path from the current state to the target location and then execute the pick-up operation after arrival.

To realize sub-goal-based action planning, we combine the LLM's natural language understanding and reasoning abilities. The procedure is as follows:
\begin{itemize}
    \item \textbf{Semantic analysis and refinement of sub-goals:} after receiving a sub-goal, the LLM first performs semantic analysis to identify the key elements in the sub-goal, such as location movement implied by ``go to $x$'' and the pick-up operation implied by ``take an item.'' The system then refines the sub-goal further according to the current game state, such as the player's current position and environment information.
    \item \textbf{Path planning and action generation:} for sub-goals that require position change, the system combines path-planning algorithms (such as A* or Dijkstra) with map information to compute an optimal path from the current state to the target state. Based on the path-planning result, it generates the corresponding action sequence, such as ``move up'' and ``move left.'' After reaching the target position, the system then generates subsequent operations such as ``pick up'' or ``use''.
    \item \textbf{Action sequence optimization:} after generating an initial action sequence, the LLM further optimizes it by removing redundant actions so that the task can be completed in as few steps as possible.
\end{itemize}

During execution, the LLM can also dynamically adjust the action sequence according to real-time environmental changes. For example, if the player encounters an obstacle, the system replans the path and generates a new action sequence online.
The LLM communicates with the game engine through an API interface. The engine sends sub-goals to the LLM, the LLM returns generated action sequences, and the engine executes them. In this way, complex game quests can be automatically decomposed into concrete execution steps, substantially improving the automation and efficiency of quest execution.

\subsection{Game State Transition Graph and Test Path Generation}
In our framework, the game environment is modeled as a dynamically updated graph structure, where nodes represent specific game states and edges represent executable actions that lead from one state to another. As the quest proceeds, the LLM agent continuously explores new states through interaction with the environment. Whenever an action is executed and causes an environmental transition, the framework records the corresponding state-action pair and updates the graph nodes and edges.

To ensure uniqueness and effectiveness of the graph structure, we encode each state into a fixed-length vector. This representation guarantees the uniqueness of each state while making similar states close in vector space, thereby facilitating the identification and handling of similar states. In this way, the framework can effectively manage a large state-action space in complex environments while continuously expanding test coverage through graph updates.

When quest execution ends, the graph structure often contains multiple candidate paths from the initial state to the quest completion state. To ensure both execution efficiency and test coverage, we use a $k$-shortest-path algorithm to generate multiple near-optimal candidate paths. When all state transition costs are set to 1, the first path corresponds to the minimum action sequence required to complete the quest, while the remaining paths provide additional alternatives with lengths close to the shortest one. The framework then removes duplicates and evaluates the coverage of these candidate paths, prioritizing those that cover more new states or edges. The final set of test paths can both quickly verify whether the quest can be completed and expand test coverage over a broader range, thereby improving defect discovery ability.
The resulting candidate paths contain task-relevant key actions and avoid redundant operations within the explored graph. Executing multiple candidates exposes additional state transitions and supports coverage-oriented testing under different completion paths.

\subsection{Prompt Design Examples}
In this study, prompt design is critical to improving LLM performance in game quest testing. To enable the LLM to accurately parse quests, understand game states, and generate suitable sub-goals and action sequences, we design a multi-level, step-by-step prompt framework, as shown in Fig. \ref{fig:prompt}. The framework mainly includes three parts: game information, prompt instructions, and output control. Fig. \ref{fig:prompt} summarizes the shared component structure of the prompts; the complete prompt templates actually used in the experiments, including the environment-specific game information for each game, are provided in Appendix~\ref{app:prompt}.

\emph{Game information:}
This part includes a brief overview and detailed background of the game. The brief overview helps the LLM quickly understand the game type and objectives, while the detailed information provides contextual support for task execution. For example, in Fig.~\ref{fig:prompt}(a) and (b), the prompt aims to guide the LLM to generate key sub-goals and goal-action planning functions from the developer's perspective. Therefore, this part provides developer-oriented task configuration and observation information. In Fig.~\ref{fig:prompt}(c), the prompt aims to generate key sub-goal sequences from the player's perspective, so it provides player-oriented game information and sub-goal details.

\emph{Prompt instructions:}
The instruction part specifies the concrete objective assigned to the LLM by the prompt. This part provides the key information required for generation and is customized according to the prompt objective. For example, in Fig.~\ref{fig:prompt}(a) and (b), the quest description is designed from the game developer's perspective to guide the LLM to generate suitable key sub-goals and goal-action planning functions for that game.

\begin{figure}[h!tp]
    \centering
    \begin{lstlisting}
{ "prompts": [
  { "id": "a", "role": "game developer",
    "title": "Task Analysis and Key Sub-goal Generation",
    "output_rule": "Please output a list, where each key sub-goal item is output in dictionary format, the keys need to have the name of the key sub-goal, purpose explanation..., and additional instructions...",
    "instruction": "Next, you need to, from a developer's perspective, create a series of general key sub-goals for this game to assist testers in completing game testing. A key sub-goal is defined as a set of game input actions, aimed at..., for example...",
    "game_info": { "Other introductions": "...",
      "Task types": "Task type A, task type B...",
      "Acceptable inputs": "Game input A, game input B, etc...",
      "Obtainable observation information": "Player status, NPC, map...",
      "Basic game information": "Game XX is a... game, players need to complete..." } },
  { "id": "b", "role": "game developer", "title": "Action Sequence Generation",
    "output_rule": "It should be noted that... in the output functions, it needs to include inputs and outputs, descriptions of the functions, etc.",
    "instruction": "Next, I will introduce a series of goal-based key sub-goals. You need to write functions to decompose them into the game's native inputs based on the description of each key sub-goal.",
    "key_sub_goal": { "purpose": "purpose", "description": "description...",
      "name": "Name of the key sub-goal" },
    "game_info": { "Other introductions": "...",
      "Task types": "Task type A, task type B...",
      "Acceptable inputs": "Game input A, game input B, etc...",
      "Obtainable observation information": "Player status, NPC, map...",
      "Basic game information": "Game XX is a... game, players need to complete..." } },
  { "id": "c", "role": "game player", "title": "LLM-based Goal Planning",
    "game_info": { "Other introductions": "...",
      "Basic game information": "[...]", "Observation information": "[...]",
      "Task content and description": "[...]",
      "Selectable key sub-goals and descriptions": "[...]" },
    "output_rule": [
      "In the output, the response format for your every step should be output in dictionary format, the keys need to include the action, the analysis of the state after the action is completed, the planning for the next step...",
      "Additionally, you also need to output:...",
      "(The rest of the prompt words about output formatting)" ],
    "instruction": [
      "Please provide a sequence of key sub-goals in order based on current observations and task information to complete the task...",
      "After each selection of a key sub-goal, you need to infer the new state...",
      "You also need to explicitly point out the subsequent planning to complete the task based on the inferred state after each step..." ] } ] }
    \end{lstlisting}
    \caption{Examples of prompt components}
    \label{fig:prompt}
\end{figure}

\emph{Output format and control:}
To ensure that the LLM output is consistent and machine-parsable, we require the output to follow a dictionary-like structured format. This structure specifies the key-value organization for each operation, ensuring that the name, purpose, and explanation of each step are represented in structured data for later parsing and analysis. For example, one field can be used for state analysis, while another is used for planning the next step.
At the end of the prompt in Fig.~\ref{fig:prompt}(c), we also require the LLM to include both analysis of the current state and planning for the next step when generating each operation. This operationalizes chain-of-thought prompting. Through step-by-step reasoning, the model unfolds the task progressively and ensures that each operation aligns with quest logic and game context. The CoT structure helps the model reflect on each generated action and verify whether the current step satisfies the task requirements, thereby reducing reasoning errors and improving the logical consistency of generated action sequences.

\section{Evaluation}

Each game quest testing experiment is conducted on tasks with multiple complexity levels. The research questions are as follows:
\begin{itemize}
    \item \textbf{RQ1:} Compared with traditional automated testing methods, how effective is LLMEGQT in verifying quest completability, improving testing efficiency (execution steps and environment interactions), and expanding predefined target coverage?
    \item \textbf{RQ2:} As the core driver of the framework, how do different large language models affect the overall performance of LLMEGQT?
\end{itemize}

\subsection{Experimental Setup}\label{sec:setup}

\subsubsection{Game Environments}

To evaluate the effectiveness and applicability of the proposed LLMEGQT framework across different quest-driven game settings, we designed and implemented three experimental environments: the Hero Game, the Kitchen Game, and Minecraft. The Hero Game is a classic linear story-driven role-playing game \cite{jie2024alanguage}, where the player acts as a hero, triggers quests with NPCs (non-player characters), and defeats specific enemies to advance the story.
The Kitchen Game is an action role-playing game (ARPG) \cite{overcooked_team17,Cai2024OvercookedPlus}, in which the player acts as a chef and manipulates character actions to improve the efficiency of food preparation. Minecraft is an open-world sandbox game in which players synthesize blocks, materials, and tools, and can build and explore freely.
The Hero Game is mainly used to test the framework's performance on relatively simple and linear stories. The Kitchen Game evaluates the framework's ability to handle action logic and task execution efficiently. Minecraft evaluates its performance in diverse, complex open-ended games.

\begin{figure}[h!tp]
    \centering
    \includegraphics[width=0.92\linewidth]{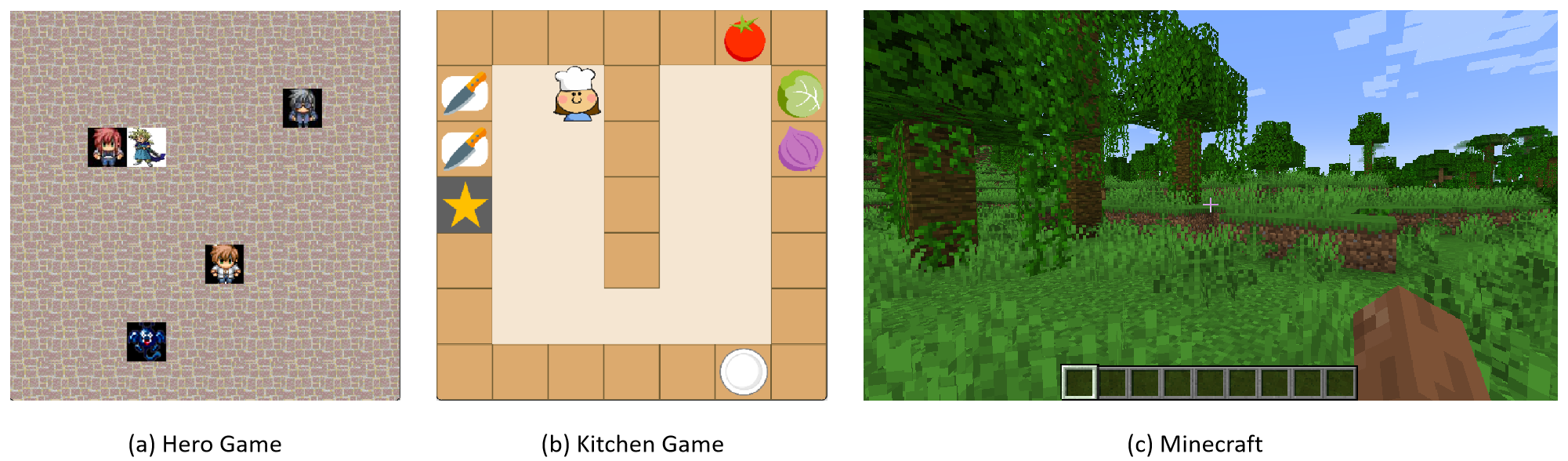}
    \caption{Screenshots of the experimental environments}
    \label{fig:gameimage}
\end{figure}

\emph{Hero Game.}
Fig. \ref{fig:gameimage}(a) shows a screenshot of the Hero Game. This environment corresponds to a classic role-playing game and is designed to evaluate the ability of each method to advance the story and navigate quests in a simple rule-based environment. To facilitate testing, we preserve the basic plot logic while simplifying the environment, so researchers can easily create customized quests. For example, a player may first need to talk to NPC1, then defeat a monster, and finally report back to NPC2.
The environment consists of four types of objects: NPCs, monsters, the player, and special terrain. The player can interact with NPCs through dialogue, item acquisition, or item submission; interactions with monsters are performed through attack commands; and special terrain can trigger specific events. The scene is defined on a fixed 10$\times$10 grid map, containing 3 NPCs, 1 monster, and 1 special terrain tile. The game state is represented as a dictionary that records all game information including the map, player, NPCs, and quest list, and the player can observe the complete environment state.

Each test quest consists of 3 to 5 stages, and the player must complete these stages in sequence to finish the quest. Each stage is defined by a quest type, a target object ID, and a textual quest description. The quest type defines the current task category, and different task types require different player actions. For example, a ``use item'' quest takes a single item ID as the target object ID, while a ``get from NPC'' quest specifies both an NPC ID and an item ID. The quest description is a short natural-language sentence that integrates the quest type and target object information in a story-like way, so as to simulate game quest narratives.
As the number of stages increases, task complexity rises gradually. In later stages, execution often depends on the results of earlier stages, which causes the state space to expand rapidly, increases the number of possible paths, and makes the quest logic more complex. At the same time, coverage also decreases as stages progress: under rapid state-space growth, testing may generate more paths in absolute terms, but the proportion of covered space becomes smaller, so potential defects are easier to miss in long quests.
The experiment contains 39 complex quests, including 16 three-stage quests, 16 four-stage quests, and 7 five-stage quests.

\emph{Kitchen Game.}
The Kitchen Game is an action simulation game in which the player aims to improve kitchen operation efficiency by controlling character actions. Its commercial prototype, the \emph{Overcooked} series, has been released on PlayStation, Xbox, Nintendo Switch, Windows, and other mainstream platforms since its debut in 2016, and has sold over 10 million copies on Steam alone.
Figure \ref{fig:gameimage}(b) shows a screenshot of the Kitchen Game. In this environment, the player must process different ingredients and eventually submit the finished dish. For example, the player may need to submit a tomato-onion salad, which requires chopping tomato and onion separately, placing them on a plate, and then submitting the plate. The Kitchen Game contains more complex rules. For example, only one ingredient can be placed on a chopping board at a time; if the wrong ingredient is placed, the player must first remove it and put it elsewhere before restarting the correct procedure. Quest completion therefore depends not only on reaching the objective but also on proper state management, requiring the player to judge the current state and infer the best path under the game rules.
Similar to the Hero Game, the Kitchen Game is also based on a grid world. The scene contains three ingredients, two chopping boards (the boards and their attached knives are immovable), one plate, and one submission counter. The plate is used to hold processed ingredients, and the submission counter is used to submit the plated dish and complete the quest. Each chopping board can process only one ingredient at a time.
In the kitchen scenario, the game state consists of map information, player information, and object information, and the player can also observe the entire game environment.
Unlike in the Hero Game, we reduce difficulty by aligning the action spaces of all compared methods with the key sub-goal space generated by LLMEGQT and then uniformly converting these sub-goals into game-action inputs. Specifically, the space contains nine goal-oriented key sub-goals such as ``(move and) pick up ingredient,'' ``go to the chopping board (and place ingredient),'' ``place the carried item on an empty table,'' ``cut,'' and ``submit.''
We design seven tasks with different difficulty levels, including three easy tasks (processing one ingredient), three medium tasks (processing two ingredients), and one difficult task (combining all ingredients). In the difficult task, because there are only two chopping boards in the scene, the player must manage resources carefully; otherwise incorrect operations will consistently increase the length of the key action sequence.

\emph{Minecraft.}
Figure \ref{fig:gameimage}(c) shows a screenshot of Minecraft.
Minecraft is an open-world sandbox game and one of the best-selling commercial games of all time. Since its release in 2009, cumulative sales have exceeded 350 million copies across PC, mobile, and all major console platforms. Its open sandbox structure and large player community have created a distinct cultural influence and also supported a diverse business ecosystem including subscriptions, content marketplaces, and educational licensing. In the game environment, players collect items through crafting and mining, use these materials to craft tools, gather more advanced items with those tools, interact with in-game entities or monsters, and build structures. For example, the quest ``kill a zombie with a stone sword'' requires players to chop wood, convert logs into planks, use planks to craft sticks, craft a wooden pickaxe, mine stone, craft a stone sword, and finally kill the zombie.
Since each step consumes resources (for example, if all sticks are consumed while crafting a wooden pickaxe, more sticks must be made before crafting a stone sword), each action must be decided according to the current state, which places high demands on state management ability.
The game state includes the player's position, inventory, and the dynamic states of spatial elements such as resource blocks, furnaces, and enemies. The player starts from a fixed position and can execute parameterized high-level commands (such as mine, craft, and attack), which are further decomposed into a series of atomic actions. The quests are based on Minecraft's built-in achievement system. We design 9 tasks divided into 3 difficulty levels: basic collection, intermediate crafting, and advanced combat, corresponding to 1-stage, 2-stage, and 3-stage completion, respectively.

\subsubsection{Evaluation Metrics}

We evaluate each method using four metrics:
\begin{itemize}
    \item \textbf{Success Rate:} The proportion of test runs in which the quest is completed successfully.
    \item \textbf{Average Steps:} The average number of interaction steps required to complete the quest, computed only over successful runs and reported as ``--'' when no run succeeds.
    \item \textbf{Environment Interactions:} The mean number of environment interaction steps consumed per run by each method during training, fitness evaluation, or LLM-guided sampling and exploration. This column uses a single unit---one interaction with the game environment---so that all methods are compared on a like-for-like basis. Higher values indicate greater time and computational cost.
    \item \textbf{Predefined Target Coverage:} The proportion of predefined state or state-action targets covered during testing.
\end{itemize}
To compare like with like, every entry in the Environment Interactions column is measured in the same unit, namely one interaction between the method and the game environment, and is reported as the mean over ten runs. For CD-PPO, each training step executes one action and therefore consumes one environment interaction. For GA, each generation evaluates every individual in the population by executing its action sequence; the reported value counts every such execution step, so it equals the mean total number of steps executed across all fitness evaluations. For LLMEGQT, no model weights are updated during testing; environment interactions are consumed by LLM-guided sampling and exploration and are counted in exactly the same way. For the Random baseline, which has no training phase, the reported value equals the mean number of rollout steps executed per run across ten runs. Under this common unit the entries are directly comparable. The run-to-run standard deviation is below 3\% of the reported mean for GA and Random and below 2\% for LLMEGQT; CD-PPO has zero variance when it uses a fixed training budget and below 5\% where early stopping applies.

Unless stated otherwise, each individual task (quest) is evaluated independently, and each task is repeated ten times; the ten runs for that task form the basis for all reported statistics. For difficulty levels containing multiple tasks (e.g., 16 three-stage quests in the Hero Game), each task is evaluated separately; the table reports results for one representative task per difficulty level, selected before any results were collected as the task whose structural complexity (number of objects and interactions) was closest to the median of that level. Average Steps is reported as the mean $\pm$ standard deviation over successful runs (``--'' when no run succeeds); Predefined Target Coverage is reported as the mean $\pm$ standard deviation across the ten runs; and Success is reported as the number of successful runs over the total number of runs (denominator always ten). The corresponding 95\% Wilson confidence intervals are summarized below.

\begin{center}
\small
\begin{tabular}{cc@{\hspace{1.6em}}cc@{\hspace{1.6em}}cc}
\toprule
\textbf{Success} & \textbf{95\% CI} & \textbf{Success} & \textbf{95\% CI} & \textbf{Success} & \textbf{95\% CI} \\
\midrule
0/10 & $[0.00, 0.28]$ & 4/10 & $[0.17, 0.69]$ & 8/10  & $[0.49, 0.94]$ \\
1/10 & $[0.02, 0.40]$ & 5/10 & $[0.24, 0.76]$ & 9/10  & $[0.60, 0.98]$ \\
2/10 & $[0.06, 0.51]$ & 6/10 & $[0.31, 0.83]$ & 10/10 & $[0.72, 1.00]$ \\
3/10 & $[0.11, 0.60]$ & 7/10 & $[0.40, 0.89]$ &       &                \\
\bottomrule
\end{tabular}
\end{center}

To evaluate LLMEGQT's ability to explore diverse paths and state spaces, we define a set of behavioral targets in each game environment. Following prior coverage-oriented game-testing studies~\cite{ariyurek2019automated,feldmeier2023fully}, each target corresponds to a task-relevant state or state-action pair along a quest-completion path. A target is covered when the corresponding state is reached or the state-action pair is executed. We instantiate three targets per quest stage from an environment-specific target schema and report the proportion of covered targets over the total target set, which serves as the same denominator for every method. Appendix~\ref{app:target} summarizes the target schemas, triggering conditions, and selection rationale. The instantiated target sets were fixed before running the experiments, were identical for all methods, and were not adjusted after any method's results were collected.

\subsubsection{Comparison Methods}

To validate the effectiveness of the proposed method, we compare LLMEGQT against three baselines: a random algorithm, a genetic algorithm (GA) \cite{holland-genetic-algorithms-1992}, and curiosity-driven PPO (CD-PPO) \cite{9619048}. We additionally evaluate LLMEGQT with five different LLM backends to analyze model-specific effects. The random algorithm completes tasks by randomly selecting available actions. GA serves as a classic search-based testing baseline. PPO is a commonly used reinforcement learning algorithm based on policy gradients, and CD-PPO augments it with intrinsic curiosity rewards to encourage the discovery of unseen game states. In our experiments, CD-PPO outputs an action probability distribution according to game observations, and a greedy strategy is then used to select the final action.

The five LLM backends are gpt-5, gpt-4.1, o4-mini, qwen3-plus, and deepseek-v3. We deliberately use general-purpose LLMs without game-domain-specific fine-tuning, rather than models specifically trained for games. This helps us understand how general LLMs perform in practice and reveals their distinct strengths and limitations. This design also provides a basis for the generality and robustness of our method across different model backends. If pretraining or fine-tuning on game-domain data becomes available, we expect further performance gains.
Detailed reproducibility settings are summarized in Appendix~\ref{app:repro}. For each LLM we report the provider, the exact API model identifier, the reasoning mode, the sampling temperature, top-$p$, the maximum output token count, and the number of samples. None of the LLM APIs used expose deterministic model snapshots or seeds; the API model identifier returned with each response is recorded in the logs. The run-level seeds and sources of stochasticity for Random, GA, CD-PPO, and LLMEGQT are also reported in Appendix~\ref{app:repro}.

The detailed parameter settings for each game environment are as follows.
In the Hero Game, the maximum number of interaction steps in each scenario is set to 5000. For the random algorithm, the upper bound for a single generated action sequence is set to 500. Each task is repeated 10 times, and the average result is reported to reduce random fluctuation. For GA, the upper bound of the individual action sequence is also 500, the population size is 50, and the maximum number of iterations is 150. For CD-PPO, we use encoded game states as input, set the maximum number of training steps to 35,000, and introduce intrinsic rewards to encourage exploration of unseen states. An early stopping strategy is used: if performance does not improve over 7 evaluation periods (5000 steps per period), training terminates early and the best model is selected for evaluation. The Hero Game reward function is $-0.1$ per step, $+10$ for a successful dialogue with the NPC designated by the current stage, $+20$ for defeating the specified monster, $+5$ for triggering the special-terrain event, $+100$ for completing a quest stage, an additional $+200$ for completing the entire quest, and $-5$ for an invalid action (e.g., using an item or attacking a target not required by the current stage). For LLMEGQT, we perform 10 sampling rounds per test, generate 5 key sub-goal sequences in each round, and generate 10 action sequences for each sub-goal. Game state information is uniformly encoded into natural language and passed to the LLM to fully exploit its language understanding and reasoning ability.
In the Kitchen Game, the maximum number of interaction steps in each scenario is set to 100, and the upper bound of a single generated action sequence for the random algorithm is also 100. The individual action sequence length for GA is 100, and the other parameters are the same as in the Hero Game. Its reward function is $-0.1$ per step, $+10$ for each processed ingredient, $+200$ for completing one dish, and $-5$ for an incorrect submission.
In Minecraft, the maximum number of interaction steps in each scenario is 500; the upper bound of a single action sequence for both the random algorithm and GA is 500, and GA uses the same population size and number of iterations; CD-PPO takes encoded state information as input, and the maximum number of training steps is 100,000. The reward function is: $-0.1$ per step, $+10$ for each key component obtained, and $+100$ for quest completion.

\subsection{RQ1: Effectiveness and Coverage}

To answer RQ1, we compare LLMEGQT against traditional baselines across all three game environments and analyze both quest completion performance and predefined target coverage.
\emph{Hero Game.}
Table \ref{tab:minigame_all} presents the results for the Hero Game. The first five rows show the results of different LLMs within the LLMEGQT framework, while the remaining rows show the comparison methods. We classify the results by the number of quest stages and report the number of successful runs, the average number of steps, and the number of environment interactions. The average number of steps is computed only over successful runs and reflects the interaction steps required to complete the quest successfully.

The results show that the random algorithm performs worst in both success rate and average number of steps, far behind the other methods. CD-PPO achieves a relatively high success rate, but its average number of steps increases substantially. GA performs reasonably well in terms of success rate, but still requires many more steps and environment interactions.
Under the LLMEGQT framework, the LLM-driven variants generally achieve higher completion rates and lower step counts than GA and CD-PPO, while requiring fewer exploration interactions. Target coverage is more mixed on easier tasks and becomes more favorable to LLMEGQT on deeper quest stages. On 3-stage quests, the Random baseline reaches a target coverage comparable to the top LLMs, because the target states of simple tasks are densely distributed in the state space and can be reached by random sampling alone; as task complexity increases, however, the Random algorithm fails to advance quest progress and its coverage drops sharply, consistent with its declining success rate.
In terms of environment interactions, GA and CD-PPO both consume large numbers of interactions, on the order of tens or hundreds of thousands. In contrast, the LLM-based methods under LLMEGQT require only a few hundred to around five hundred interactions, showing much higher efficiency.

\begin{table}[h!tp]
    \centering
    \caption{Experimental results of different methods in the Hero Game environment}
    \label{tab:minigame_all}
    \scriptsize
    \setlength{\tabcolsep}{4pt}
    \renewcommand{\arraystretch}{1.12}
    \begin{tabular}{@{}lcccc@{}}
    \toprule
    \textbf{Method} & \textbf{Success} & \textbf{Steps} & \shortstack{\textbf{Env.}\\\textbf{interactions}} & \shortstack{\textbf{Target}\\\textbf{coverage}} \\
    \midrule
    \multicolumn{5}{@{}l}{\textit{3 stages}} \\
    gpt-5          & 10/10 & 58.10$\pm$3.2  & 430.00    & 0.78$\pm$0.06 \\
    gpt-4.1        & 10/10 & 58.80$\pm$3.5  & 444.00    & 0.89$\pm$0.04 \\
    o4-mini        & 10/10 & 62.27$\pm$4.1  & 343.00    & 0.78$\pm$0.06 \\
    deepseek-v3    & 10/10 & 61.98$\pm$4.3  & 343.00    & 0.67$\pm$0.07 \\
    qwen3-plus     & 10/10 & 65.70$\pm$4.6  & 397.00    & 0.67$\pm$0.07 \\
    CD-PPO         & 9/10  & 219.8$\pm$28.5 & 35000.00  & 0.78$\pm$0.06 \\
    GA             & 10/10 & 227.0$\pm$30.1 & 189358.97 & 0.78$\pm$0.06 \\
    Random         & 7/10  & 500.0$\pm$0.0  & 500.00    & 0.89$\pm$0.05 \\
    \addlinespace
    \multicolumn{5}{@{}l}{\textit{4 stages}} \\
    gpt-5          & 10/10 & 72.10$\pm$4.1  & 520.00    & 0.83$\pm$0.05 \\
    gpt-4.1        & 10/10 & 75.60$\pm$4.5  & 539.00    & 0.78$\pm$0.06 \\
    o4-mini        & 9/10  & 79.42$\pm$5.6  & 428.00    & 0.78$\pm$0.06 \\
    deepseek-v3    & 9/10  & 80.23$\pm$5.9  & 428.00    & 0.83$\pm$0.05 \\
    qwen3-plus     & 10/10 & 76.31$\pm$5.2  & 507.00    & 0.83$\pm$0.05 \\
    CD-PPO         & 4/10  & 299.7$\pm$35.1 & 35000.00  & 0.72$\pm$0.07 \\
    GA             & 7/10  & 279.5$\pm$33.8 & 238302.57 & 0.67$\pm$0.07 \\
    Random         & 1/10  & 500.0$\pm$0.0  & 500.00    & 0.22$\pm$0.06 \\
    \addlinespace
    \multicolumn{5}{@{}l}{\textit{5 stages}} \\
    gpt-5          & 10/10 & 84.50$\pm$5.3  & 560.00    & 0.52$\pm$0.07 \\
    gpt-4.1        & 10/10 & 85.70$\pm$5.1  & 580.00    & 0.44$\pm$0.08 \\
    o4-mini        & 9/10  & 87.52$\pm$6.2  & 534.00    & 0.37$\pm$0.07 \\
    deepseek-v3    & 6/10  & 90.61$\pm$6.8  & 523.00    & 0.44$\pm$0.08 \\
    qwen3-plus     & 7/10  & 88.29$\pm$6.5  & 571.00    & 0.44$\pm$0.08 \\
    CD-PPO         & 3/10  & 345.8$\pm$41.2 & 35000.00  & 0.22$\pm$0.06 \\
    GA             & 6/10  & 317.1$\pm$38.6 & 270443.92 & 0.15$\pm$0.05 \\
    Random         & 0/10  & --             & 500.00    & 0.04$\pm$0.03 \\
    \bottomrule
    \end{tabular}
\end{table}

\emph{Kitchen Game.}
The results are shown in Table \ref{tab:overcooked_all}. The methods show clear descriptive differences on easy and medium tasks, and the gap becomes more apparent on difficult tasks.
On easy tasks, except for the random algorithm, all methods achieve a 100\% success rate.
On medium tasks, as task difficulty increases, GA and CD-PPO show lower success rates and require more steps, while the random algorithm cannot complete the tasks. The LLM-based framework still performs well in step optimization, maintaining high success rates while requiring relatively few steps.
On difficult tasks, traditional methods (CD-PPO, random, and GA) all fail to complete the quest, showing their limitations on high-complexity tasks. Under LLMEGQT, the LLM-driven methods still achieve partial to full completion, demonstrating the clear advantage of the framework on complex tasks.

As in the Hero Game, LLMEGQT consumes fewer environment interactions, showing higher efficiency and lower resource consumption.
In terms of coverage, all LLMs reach relatively high levels on easy and medium tasks. On difficult tasks, coverage decreases across the board, reflecting the increased difficulty of maintaining high coverage on harder quests. The detailed comparison among different LLMs is presented in RQ2.

\begin{table}[h!tp]
    \centering
    \caption{Experimental results of different methods in the Kitchen Game environment}
    \label{tab:overcooked_all}
    \scriptsize
    \setlength{\tabcolsep}{4pt}
    \renewcommand{\arraystretch}{1.12}
    \begin{tabular}{@{}lcccc@{}}
    \toprule
    \textbf{Method} & \textbf{Success} & \textbf{Steps} & \shortstack{\textbf{Env.}\\\textbf{interactions}} & \shortstack{\textbf{Target}\\\textbf{coverage}} \\
    \midrule
    \multicolumn{5}{@{}l}{\textit{Easy}} \\
    gpt-5          & 10/10 & 6.40$\pm$1.1   & 340.00    & 0.89$\pm$0.04 \\
    gpt-4.1        & 10/10 & 6.61$\pm$1.2   & 352.00    & 0.89$\pm$0.04 \\
    o4-mini        & 10/10 & 6.73$\pm$1.3   & 326.00    & 0.67$\pm$0.07 \\
    deepseek-v3    & 10/10 & 6.76$\pm$1.4   & 336.00    & 0.78$\pm$0.06 \\
    qwen3-plus     & 10/10 & 6.81$\pm$1.3   & 326.00    & 0.78$\pm$0.06 \\
    CD-PPO         & 10/10 & 6.00$\pm$1.0   & 35000.00  & 0.89$\pm$0.04 \\
    GA             & 10/10 & 22.00$\pm$5.4  & 178486.67 & 0.89$\pm$0.04 \\
    Random         & 7/10  & 99.33$\pm$12.5 & 99.53     & 0.89$\pm$0.04 \\
    \addlinespace
    \multicolumn{5}{@{}l}{\textit{Medium}} \\
    gpt-5          & 10/10 & 11.20$\pm$1.8  & 560.00    & 0.89$\pm$0.04 \\
    gpt-4.1        & 10/10 & 11.65$\pm$1.9  & 576.00    & 0.78$\pm$0.06 \\
    o4-mini        & 10/10 & 11.18$\pm$1.7  & 561.00    & 0.72$\pm$0.07 \\
    deepseek-v3    & 9/10  & 11.09$\pm$1.8  & 569.00    & 0.78$\pm$0.06 \\
    qwen3-plus     & 10/10 & 11.57$\pm$1.9  & 609.00    & 0.72$\pm$0.07 \\
    CD-PPO         & 4/10  & 70.67$\pm$18.3 & 35000.00  & 0.67$\pm$0.07 \\
    GA             & 7/10  & 40.67$\pm$8.9  & 181220.67 & 0.78$\pm$0.06 \\
    Random         & 0/10  & --              & 100.00    & 0.17$\pm$0.05 \\
    \addlinespace
    \multicolumn{5}{@{}l}{\textit{Hard}} \\
    gpt-5          & 10/10 & 16.00$\pm$2.3  & 782.00    & 0.67$\pm$0.07 \\
    gpt-4.1        & 9/10  & 16.00$\pm$2.5  & 778.00    & 0.56$\pm$0.08 \\
    o4-mini        & 10/10 & 17.50$\pm$2.8  & 804.00    & 0.56$\pm$0.08 \\
    deepseek-v3    & 5/10  & 16.76$\pm$2.9  & 676.00    & 0.56$\pm$0.08 \\
    qwen3-plus     & 6/10  & 16.36$\pm$2.6  & 493.00    & 0.67$\pm$0.07 \\
    CD-PPO         & 0/10  & --              & 35000.00  & 0.22$\pm$0.06 \\
    GA             & 0/10  & --              & 173603.00 & 0.11$\pm$0.04 \\
    Random         & 0/10  & --              & 100.00    & 0.11$\pm$0.04 \\
    \bottomrule
    \end{tabular}
\end{table}

\emph{Minecraft.}
Table \ref{tab:mc_all} shows the experimental results in Minecraft. Overall, the LLMEGQT framework performs favorably in completion rate, step optimization, environment interactions, and target coverage.
On easy tasks, all LLMs achieve a 100\% completion rate with optimized step counts, while CD-PPO and GA also achieve full completion but require more steps and incur much higher environment-interaction costs.
On medium tasks, the completion rate of traditional methods declines, and the random algorithm drops to 0. The LLM-based framework maintains high completion rates and step efficiency, while traditional methods require consistently more steps and show lower target coverage.
On hard tasks, all traditional methods fail to complete the quest, while LLMEGQT remains effective, demonstrating the clear advantage of LLM-driven testing on high-difficulty tasks.
In terms of environment interactions, LLMEGQT remains at a low level across all difficulty levels, while the traditional methods require tens of thousands to hundreds of thousands of interactions. The detailed comparison among different LLMs is presented in RQ2.

\begin{table}[h!tp]
    \centering
    \caption{Experimental results of different methods in the Minecraft environment}
    \label{tab:mc_all}
    \scriptsize
    \setlength{\tabcolsep}{4pt}
    \renewcommand{\arraystretch}{1.12}
    \begin{tabular}{@{}lcccc@{}}
    \toprule
    \textbf{Method} & \textbf{Success} & \textbf{Steps} & \shortstack{\textbf{Env.}\\\textbf{interactions}} & \shortstack{\textbf{Target}\\\textbf{coverage}} \\
    \midrule
    \multicolumn{5}{@{}l}{\textit{1 stage}} \\
    gpt-5          & 10/10 & 4.00$\pm$0.0  & 600.00    & 0.89$\pm$0.04 \\
    gpt-4.1        & 10/10 & 4.00$\pm$0.0  & 599.00    & 0.78$\pm$0.06 \\
    o4-mini        & 10/10 & 4.33$\pm$0.5  & 697.00    & 0.78$\pm$0.06 \\
    deepseek-v3    & 10/10 & 4.00$\pm$0.0  & 600.00    & 0.56$\pm$0.08 \\
    qwen3-plus     & 10/10 & 4.07$\pm$0.3  & 601.00    & 0.78$\pm$0.06 \\
    CD-PPO         & 10/10 & 6.67$\pm$1.6  & 32730.05  & 0.89$\pm$0.04 \\
    GA             & 10/10 & 6.00$\pm$1.4  & 451877.67 & 0.78$\pm$0.06 \\
    Random         & 7/10  & 387.3$\pm$58.6 & 421.11    & 0.89$\pm$0.04 \\
    \addlinespace
    \multicolumn{5}{@{}l}{\textit{2 stages}} \\
    gpt-5          & 10/10 & 6.17$\pm$1.2  & 1052.00   & 0.83$\pm$0.05 \\
    gpt-4.1        & 10/10 & 5.67$\pm$1.1  & 950.00    & 0.83$\pm$0.05 \\
    o4-mini        & 10/10 & 6.77$\pm$1.4  & 1149.00   & 0.67$\pm$0.07 \\
    deepseek-v3    & 10/10 & 6.71$\pm$1.3  & 969.00    & 0.78$\pm$0.06 \\
    qwen3-plus     & 8/10  & 6.30$\pm$1.5  & 855.00    & 0.72$\pm$0.07 \\
    CD-PPO         & 10/10 & 14.00$\pm$3.2 & 47752.65  & 0.56$\pm$0.08 \\
    GA             & 7/10  & 10.00$\pm$2.3 & 566419.00 & 0.61$\pm$0.07 \\
    Random         & 0/10  & --             & 500.00    & 0.11$\pm$0.04 \\
    \addlinespace
    \multicolumn{5}{@{}l}{\textit{3 stages}} \\
    gpt-5          & 10/10 & 9.60$\pm$1.8  & 1542.00   & 0.56$\pm$0.08 \\
    gpt-4.1        & 10/10 & 9.37$\pm$1.9  & 1501.00   & 0.52$\pm$0.07 \\
    o4-mini        & 9/10  & 9.80$\pm$2.1  & 1693.00   & 0.37$\pm$0.07 \\
    deepseek-v3    & 10/10 & 9.04$\pm$1.7  & 1242.00   & 0.44$\pm$0.08 \\
    qwen3-plus     & 6/10  & 10.83$\pm$2.4 & 955.00    & 0.44$\pm$0.08 \\
    CD-PPO         & 0/10  & --             & 100000.00 & 0.07$\pm$0.03 \\
    GA             & 0/10  & --             & 735810.67 & 0.11$\pm$0.04 \\
    Random         & 0/10  & --             & 500.00    & 0.00$\pm$0.0 \\
    \bottomrule
    \end{tabular}
\end{table}

\emph{Cross-Game Comparison.}
Across all three environments, the LLMEGQT framework shows substantially better performance than traditional algorithms on complex and multi-stage tasks. Traditional methods such as the random algorithm can achieve some success on simple tasks, but because they cannot effectively exploit environment information, they show clear limitations on complex quests.
Similarly, although GA performs better than the random algorithm in success rate and average number of steps, it requires many environment interactions during training for shortest-path discovery, and the computational cost rises substantially as problem complexity increases.
Curiosity-driven CD-PPO performs better than GA overall in success rate and average number of steps, but is still constrained by the many environment interactions required during training and by the higher computational cost of the algorithm itself.

In contrast, the LLM-based LLMEGQT framework leverages the prior knowledge encoded in LLMs to perform effective reasoning and generate reliable paths, improving quest completion efficiency. Because LLMEGQT advances farther through the quest, it can also cover more later-stage targets. The graph's path-consolidation role is discussed further in the Discussion section.

\begin{figure}[h!tp]
    \centering
    \includegraphics[width=0.92\linewidth]{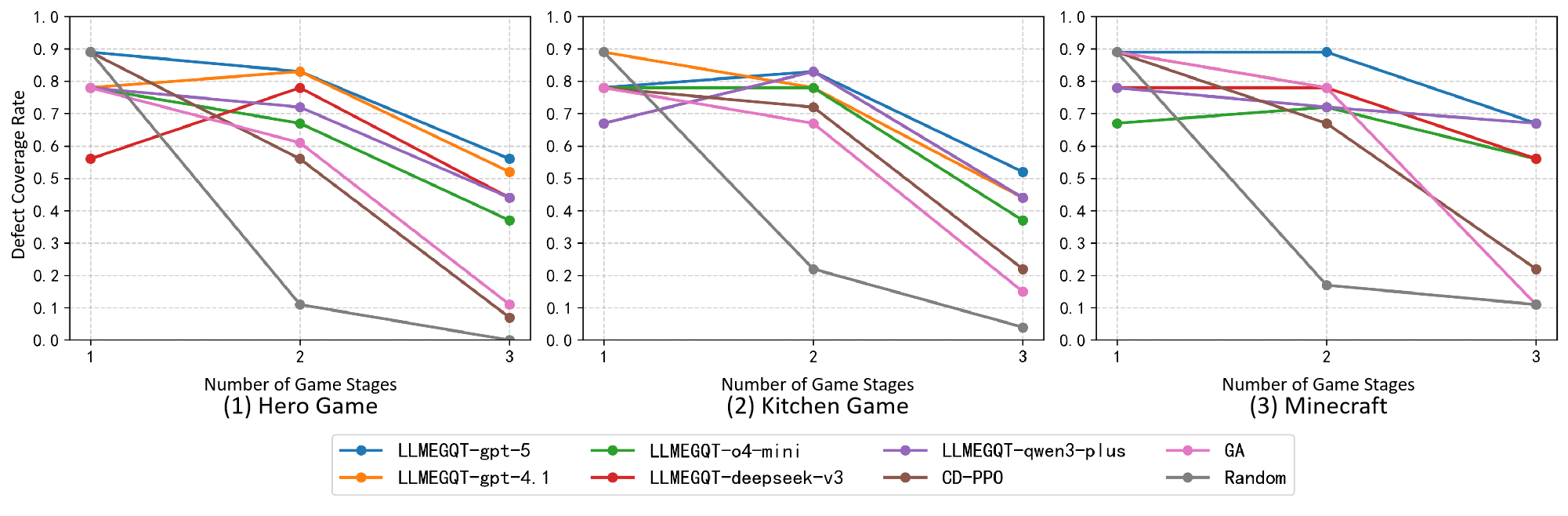}
    \caption{Comparison of predefined target coverage across experimental environments}
    \label{fig:space_stage}
\end{figure}

\begin{figure}[h!tp]
    \centering
    \includegraphics[width=0.64\linewidth]{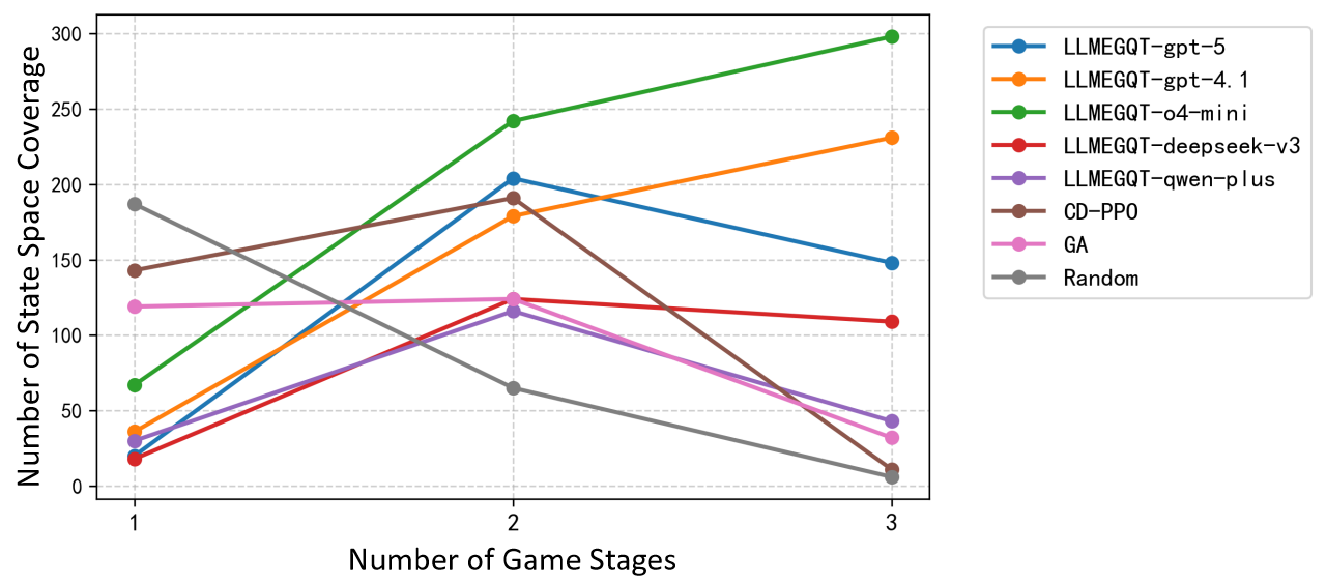}
    \caption{Comparison of the number of traversed states of different methods in Minecraft}
    \label{fig:space_count_stage_mc}
\end{figure}

\emph{State-Space Exploration and Target Coverage.}
Coverage of task-relevant behavior depends heavily on the state-space search ability of the testing algorithm \cite{zheng2019wuji}. Broad search near the initial state alone is insufficient: game quest testing must also advance quest progress so that targets in deeper stages can be reached.
Traditional state-space search methods such as the random algorithm may cover many states at the beginning, but they are inefficient at advancing the quest and therefore become limited in later stages. The random algorithm performs relatively well early on, but without effective guidance, its exploration range shrinks rapidly as the quest progresses. In contrast, methods that can effectively advance the quest can both cover the state space broadly and reach deeper stages of the game, thus discovering more latent defects.
To quantify state-space exploration and quest progression, we relate the number of predefined targets covered to game progress for LLMEGQT (driven by gpt-5), CD-PPO, GA, and the random algorithm, as shown in Fig. \ref{fig:space_stage}. We also count the total number of traversed states for each algorithm while attempting to complete tasks in Minecraft, as shown in Fig. \ref{fig:space_count_stage_mc}.

The results show that although the random algorithm explores many states in the early quest stages, it cannot advance the quest flow and therefore performs almost no further exploration in later stages. GA performs better than the random algorithm in the late stages, but because its overall quest completion success rate is low, it still fails to explore later state space effectively. In addition, both methods are weaker than the random algorithm in the initial stage, which further limits their overall exploration capability.
By contrast, LLMEGQT shows clear advantages in quest advancement. Using fewer time steps than the traditional algorithms, it reaches more states and predefined targets in later quest stages. Notably, gpt-5 has lower state coverage than gpt-4.1 and o4-mini but higher target coverage, suggesting that broad state exploration does not necessarily maximize task-relevant coverage. The detailed differences among LLMs are discussed in RQ2.

Overall, an effective game testing method should balance quest progression and state-space exploration. The LLMEGQT framework shows clear advantages in both aspects. Compared with traditional algorithms, LLMEGQT can cover more states within fewer time steps and maintain relatively high exploration efficiency even in the later quest stages.

\textit{Summary of RQ1.}
LLMEGQT generally achieves higher quest completion rates and better step efficiency than the traditional baselines, particularly on medium and difficult tasks. Target coverage is comparable or occasionally lower on some easy tasks, but LLMEGQT reaches more later-stage targets as quest depth increases. The framework also uses substantially fewer exploration interactions than the training-based baselines.

\subsection{RQ2: Impact of Different LLMs}

To answer RQ2, we analyze the performance differences among the five LLMs used to drive LLMEGQT.
The experimental results reveal clear differences among large language models within the LLMEGQT framework, especially in success rate and average number of steps. gpt-5 and gpt-4.1 perform best on most tasks, both reaching 100\% success rates and average step counts close to the theoretical optimum. Although other models such as o4-mini, qwen3-plus, and deepseek-v3 do not always match gpt-5 or gpt-4.1, they still consistently outperform traditional algorithms. Overall, the effectiveness of LLMEGQT is closely related to the capability of the underlying LLM \cite{huggingface_openllm_2024}.

As shown in Tables \ref{tab:minigame_all}--\ref{tab:mc_all}, the performance gap among LLMs becomes more pronounced as task difficulty increases.
In the Hero Game (Table \ref{tab:minigame_all}), gpt-5 achieves a 100\% success rate at all difficulty levels with the lowest average steps (58.10, 72.10, 84.50 for 3-, 4-, 5-stage quests), indicating the strongest strategy generation ability. gpt-4.1 follows closely with 100\% success but slightly more steps. o4-mini maintains a 100\% success rate on 3-stage quests but drops to 90\% on 4- and 5-stage quests. deepseek-v3 shows a notable decline to 60\% on 5-stage quests, while qwen3-plus reaches 70\%.
In the Kitchen Game (Table \ref{tab:overcooked_all}), on easy tasks all LLMs achieve 100\% success, with gpt-5 demonstrating the best step efficiency (6.40 steps). On medium tasks, gpt-5 and o4-mini maintain 100\% success while deepseek-v3 drops to 90\%. On difficult tasks, gpt-5 maintains a 100\% completion rate with the fewest steps (16.00); o4-mini also achieves 100\%; gpt-4.1 reaches 90\%; while deepseek-v3 and qwen3-plus achieve 50\% and 60\%, respectively. For target coverage, gpt-5 reaches 0.89 on both easy and medium tasks; gpt-4.1 matches 0.89 on easy tasks and reaches 0.78 on medium tasks. On difficult tasks, gpt-5 and qwen3-plus both reach 0.67.
In Minecraft (Table \ref{tab:mc_all}), gpt-5 and gpt-4.1 both maintain 100\% completion rates across all difficulty levels, with gpt-5 requiring 4.00, 6.17, and 9.60 steps respectively. o4-mini achieves 90\% on hard tasks with 9.80 steps. deepseek-v3 achieves 100\% completion on hard tasks with slightly fewer steps than gpt-5 (9.04 vs.\ 9.60), while qwen3-plus drops to 60\% on hard tasks. In terms of target coverage, gpt-5 achieves the highest values (0.89, 0.83, and 0.56 across difficulty levels), with ties on some tasks.
No single LLM minimizes exploration interactions in every setting: o4-mini and deepseek-v3 use fewer interactions in several Hero Game configurations, while qwen3-plus is lowest on some Kitchen Game and Minecraft tasks.

The relative ordering of qwen3-plus and deepseek-v3 varies by environment and metric. qwen3-plus has higher completion rates on some Hero Game and Kitchen Game tasks, whereas deepseek-v3 is stronger on the hardest Minecraft task. This variation may reflect differences in prompt composition and reasoning demands, but the current experiment does not isolate their cause.

In terms of state-space exploration (Fig. \ref{fig:space_stage} and Fig. \ref{fig:space_count_stage_mc}), the differences among LLMs are also notable. Although deepseek-v3 and qwen3-plus are slightly weaker than gpt-5, gpt-4.1, and o4-mini on some tasks, their exploration remains competitive. Interestingly, gpt-5 has lower state coverage than gpt-4.1 and o4-mini yet higher predefined target coverage, suggesting that the relationship between broad exploration and task-relevant coverage is not strictly monotonic.

To further probe the limits of LLM-driven testing, we conduct a stress test in the Minecraft environment. Specifically, we use gpt-5, which performs best in the previous experiments, to drive LLMEGQT on longer multi-stage quests. The most difficult task is ``kill a zombified piglin.'' Unlike the previously defined one-stage to three-stage tasks, this quest has ten stages: obtain planks, obtain a wooden pickaxe, obtain a stone pickaxe, obtain an iron pickaxe, obtain an iron sword, obtain a diamond pickaxe, mine obsidian, build a Nether portal, enter the Nether, and find and defeat the zombified piglin.
The results are shown in Table \ref{tab:pressure}. The completion rate of gpt-5 begins to decline from stage 6 and drops to 0 by stage 10, because such a long multi-stage task places extremely high demands on the model's planning and reasoning ability.

\begin{table}[h!tp]
\centering
\caption{Task completion rates at different stages in the stress test}
\label{tab:pressure}
\begin{tabular}{|c|c|c|c|c|c|c|}
\hline
\textbf{Stage 4} & \textbf{Stage 5} & \textbf{Stage 6} & \textbf{Stage 7} & \textbf{Stage 8} & \textbf{Stage 9} & \textbf{Stage 10} \\
\hline
1.00 & 1.00 & 0.90 & 0.80 & 0.60 & 0.30 & 0.00 \\
\hline
\end{tabular}
\end{table}

\textit{Summary of RQ2.}
The effectiveness of LLMEGQT is closely tied to the capability of the underlying LLM. gpt-5 and gpt-4.1 consistently achieve the highest success rates and step efficiency. Differences among models vary by game environment, suggesting that prompt composition and reasoning task type influence model-specific performance. The stress test reveals that even the best-performing model faces limitations on very long multi-stage quests, indicating room for future improvement.

\subsection{Discussion and Limitations}

The experimental results across three game environments reveal several high-level insights. First, replacing training-based exploration with LLM-guided sampling reduces environment interactions by roughly one to three orders of magnitude, depending on the baseline and task. Second, LLM reasoning and graph-based planning play complementary roles: the LLM provides semantic decomposition, while the graph records and optimizes explored transitions. Third, results across a linear RPG, a cooking simulation, and an open-world sandbox provide initial evidence of applicability across different quest structures, rather than comprehensive generalizability.

\emph{Illustrative graph-based path consolidation.}

Table \ref{tab:hard2} presents an illustrative difficult-task case in the Kitchen Game. Across 10 raw-LLM inference attempts, deepseek-v3 produces two successful runs with an average of 19 steps. The graph built from multi-round exploration contains five feasible completion paths, including a 16-step shortest path. These quantities describe different units---successful inference runs and feasible graph paths---and therefore are not interpreted as a controlled success-rate comparison.

\begin{table}[h!tp]
\centering
\caption{Illustrative raw-LLM and graph path results for a difficult Kitchen Game task}
\label{tab:hard2}
\begin{tabularx}{\textwidth}{|>{\centering\arraybackslash}X|>{\centering\arraybackslash}X|>{\centering\arraybackslash}X|>{\centering\arraybackslash}X|}
\hline
\textbf{Successful Raw-LLM Runs (of 10)} & \textbf{Avg. Raw-LLM Steps} & \textbf{Graph Shortest Path} & \textbf{Feasible Graph Paths} \\
\hline
2 & 19 & 16 & 5 \\
\hline
\end{tabularx}
\end{table}

The graph records explored state transitions and permits path reuse and optimization. In the illustrative Kitchen Game case, it recovers five feasible paths, including a 16-step path that is shorter than the 19-step average of the two successful raw-LLM runs. Because this case does not compare matched repetitions with and without the graph, it supports the graph's path-consolidation role but not a causal claim about success-rate improvement. A matched graph ablation across all tasks would be required to quantify a causal effect on completion rate.

Although the LLMEGQT framework demonstrates high effectiveness in our experiments, its applicability still has several limitations.
First, in the Hero Game, the quest narratives involved in the experiments are relatively simple and mainly depend on basic game interaction and path-planning logic. The underlying logic of most role-playing games comes from broadly accepted real-world commonsense and cultural background (for example, martial-arts culture), which are typically well represented in the pretraining data of general-purpose LLMs. Therefore, LLMEGQT can effectively parse and apply such commonsense knowledge in these common scenarios. However, in role-playing games with more complex plots or deeper reasoning requirements, especially those with unique world-building and rich background knowledge, the quest logic may go beyond ordinary commonsense. For example, detective games may require deeper contextual analysis and logical reasoning, or long-term tracking and understanding of complex factors such as character relations and event order. In these scenarios, the reasoning ability of general-purpose LLMs may become a bottleneck. To improve LLMEGQT in such settings, one promising direction is to fine-tune models on texts specific to the target game world.

Second, in the Kitchen Game experiments, the study assumes a single-player perspective and does not consider the complexity of cooperative multiplayer tasks. If multiplayer cooperation is introduced, the complexity of both states and actions increases substantially, especially in environments requiring coordination among multiple participants. The current framework does not test such cooperative logic. Future research should improve LLMEGQT's state-reasoning and action-optimization capabilities in collaborative tasks so that it can better handle multiplayer interaction and coordination challenges.

In addition, the scalability of the state transition graph remains limited. As the game state space, quest complexity, or action sequence length increases, the graph size can grow substantially, leading to higher costs for graph storage, updates, and shortest-path computation. In extremely complex multi-stage tasks or multiplayer cooperative tasks, the graph may grow to tens of thousands of nodes or more, at which point the efficiency of graph construction, updating, and traversal may degrade. To alleviate this issue, one possible direction is to introduce graph compression and optimize graph operations, thereby improving the scalability and efficiency of large-scale state transition graphs.

The experiments also omit strategy role-playing games because suitable open-source examples were unavailable. Strategy RPGs are characterized by decision making based on numerical trade-offs and strategic gameplay. Given the research potential already shown by LLMs in game theory \cite{fan2024can}, LLMEGQT may also apply to strategy role-playing games. More broadly, extending LLMEGQT to other game genres, especially those that can be modeled as MDPs such as card games and simulation games, is also a worthwhile research direction.

Finally, beyond game quest testing, LLMEGQT can be viewed as a framework for text-guided, sequential, and stateful interactive environments. The selected quest-driven environments combine textual instructions, long-horizon dependencies, dynamic state changes, and multiple feasible execution paths. Similar ideas could extend to workflow-oriented software testing, task-driven agent evaluation, and simulation-based validation when suitable domain-specific action and state abstractions are available.

\section{Conclusion and Future Work}
This paper proposes LLMEGQT, an LLM-based automated game quest testing framework. By combining online planning with the natural language understanding and reasoning ability of LLMs, the framework can parse quest descriptions, decompose sub-goals, and generate effective action sequences, thereby enabling automated testing of complex game quests.
We conduct experiments in three game environments, and the results show that LLMEGQT performs well across different game scenarios, especially on complex and multi-stage tasks. Traditional random and genetic algorithms have difficulty advancing through deeper quest stages and require substantially more environment interactions. Performance also varies with the underlying LLM. LLMEGQT combines early-stage exploration with effective quest advancement, leading to stronger coverage of predefined targets in later stages.

Several directions remain for further optimization and extension to improve practical impact and applicability. First, LLMEGQT could be combined with search-based methods such as heuristic search to increase breadth while maintaining depth, so that more states can be traversed in complex tasks and more potential defects can be discovered. Second, the LLM and graph structures could be integrated more deeply. Their current interaction mainly depends on sub-goal sequences generated by the LLM; future work could feed graph information directly back into the LLM to improve path generation quality and guide generation according to the currently explored graph. Third, LLMEGQT could be extended from functional testing to regression testing. By recording and analyzing previous quest paths and states, LLMEGQT could quickly generate new test cases to check whether previously fixed problems reappear in new versions, thereby reducing manual testing effort and improving coverage.

\section*{Acknowledgements}
This work was conducted in part during the first author's internship at Waseda University.

\section*{Statements and Declarations}

\subsection*{Funding}
This work was supported in part by the Japan Society for the Promotion of Science (JSPS) KAKENHI under Grant No.~25K15290.

\subsection*{Competing Interests}
The authors have no relevant financial or non-financial interests to disclose.

\subsection*{Ethics Approval}
Not applicable.

\subsection*{Consent to Participate}
Not applicable.

\subsection*{Consent for Publication}
Not applicable.

\subsection*{Author Contributions}
Enhong Mu: Methodology, software, investigation, data curation, formal analysis, visualization, and writing -- original draft. Nianyu Li and Jinyu Cai: Methodology development, technical validation, and writing -- review and editing. Mingyue Zhang, Kenji Tei, and Jialong Li: Conceptualization, supervision, project administration, and writing -- review and editing. Jialong Li: Funding acquisition. All authors read and approved the final manuscript.

\subsection*{Data Availability}
The data and code supporting the findings of this study are available from the corresponding author upon reasonable request.

\begin{CJK*}{UTF8}{gbsn}
\bibliographystyle{sn-basic}
\bibliography{bib}
\end{CJK*}

\appendix
\section{Prompt Templates}\label{app:prompt}

The framework uses three prompting tasks: (a) developer-oriented task analysis and key sub-goal generation, (b) developer-oriented action sequence generation, and (c) player-oriented online goal planning. Fig.~\ref{fig:prompt} presents the shared component structure of the prompts. Below we provide the complete prompt templates actually used in the experiments for the Hero Game. The Kitchen Game and Minecraft templates follow the same structure with environment-specific \emph{game\_info} fields and are not reproduced in this appendix.

Listing~\ref{lst:prompt_analysis} shows the complete developer-oriented task analysis and key sub-goal generation prompt, Listing~\ref{lst:prompt_action} shows the complete action sequence generation prompt, and Listing~\ref{lst:prompt_goal} shows the complete player-oriented goal-planning prompt. All templates are sent to the LLM via the API with the game state encoded as natural language.

\begin{lstlisting}[caption={Complete developer-oriented task analysis and key sub-goal generation prompt (Hero Game)},label={lst:prompt_analysis}]
{
  "prompt": {
    "id": "hero_task_analysis",
    "role": "game developer",
    "game_info": {
      "basic_game_information": "A text-based role-playing game on a 10x10 grid. The player interacts with NPCs, monsters, and special terrain to complete multi-stage quests.",
      "task_types": "talk, attack, use_item, submit_item, terrain_event",
      "quest_description": "A multi-stage quest description, e.g., 'Talk to NPC1, defeat the forest monster, and report back to NPC2.'",
      "stage_count": "The total number of stages in the quest."
    },
    "output_rule": [
      "Respond with a JSON array, one entry per stage.",
      "Each entry must include: stage_index (int), stage_type (string), target_object (string), key_sub_goals (list of strings), and dependencies (list of stage indices this stage depends on).",
      "Key sub-goals must be atomic and directly realizable by the game action space."
    ],
    "instruction": [
      "Analyze the quest description and identify each stage's type, target, and required operations.",
      "For each stage, enumerate the key sub-goals that must be completed in order.",
      "Identify dependencies between stages that constrain execution order."
    ]
  }
}
\end{lstlisting}

\begin{lstlisting}[caption={Complete player-oriented goal-planning prompt (Hero Game)},label={lst:prompt_goal}]
{
  "prompt": {
    "id": "hero_goal_planning",
    "role": "game player",
    "game_info": {
      "basic_game_information": "You are a hero in a text-based role-playing game on a 10x10 grid. You can move up/down/left/right, talk to NPCs, attack monsters, use or submit items, and step on special terrain. You must complete the quest by satisfying every stage in order.",
      "observation_information": "You receive the full game state as a dictionary: player position and status, NPC positions and dialogue states, monster positions, special terrain, inventory, and the current quest list.",
      "task_content_and_description": "The quest description of the current stage, e.g., 'Talk to NPC1, defeat the forest monster, and report back to NPC2.'",
      "selectable_key_sub_goals_and_descriptions": [
        "move to (x,y) and talk to an NPC",
        "move to (x,y) and attack a monster",
        "use an item from the inventory",
        "submit an item to an NPC",
        "step on a special terrain tile"
      ]
    },
    "output_rule": [
      "Respond with exactly one JSON object per step: {\"sub_goal\": ..., \"state_reasoning\": ..., \"next_planning\": ...}.",
      "The sub_goal must be selected from the provided selectable key sub-goals.",
      "Continue step by step until the quest is completed.",
      "Do not repeat any sub-goal sequence proposed in previous rounds."
    ],
    "instruction": [
      "Analyze the current observation and the quest description.",
      "Select the next key sub-goal that advances the quest toward completion.",
      "Reason about the expected state after executing the sub-goal.",
      "Plan the subsequent sub-goal based on the inferred state."
    ]
  }
}
\end{lstlisting}

\begin{lstlisting}[caption={Complete action sequence generation prompt (Hero Game)},label={lst:prompt_action}]
{
  "prompt": {
    "id": "hero_action_planning",
    "role": "game developer",
    "game_info": {
      "basic_game_information": "A text-based role-playing game on a 10x10 grid in which the player interacts with NPCs, monsters, and special terrain.",
      "task_types": "talk, attack, use_item, submit_item, terrain_event",
      "acceptable_inputs": [
        "move_up", "move_down", "move_left", "move_right",
        "talk(npc_id)", "attack(monster_id)", "use(item_id)",
        "submit(npc_id, item_id)", "step_on_terrain(terrain_id)"
      ],
      "obtainable_observation_information": "player status, NPC states, monster states, inventory, map"
    },
    "output_rule": "Output a list of JSON functions. Each function must contain its name, its input/output types, and a description of how it decomposes the key sub-goal into native game inputs.",
    "instruction": "I will give you a series of goal-based key sub-goals. For each sub-goal, write a function that decomposes it into the game's native inputs.",
    "key_sub_goal": {
      "name": "Name of the key sub-goal",
      "purpose": "Purpose of the key sub-goal",
      "description": "Description of the key sub-goal"
    }
  }
}
\end{lstlisting}

\section{Predefined Target Schemas}\label{app:target}

The target sets were fixed before the experiments, were identical for all methods, and every method was evaluated against the same denominator, namely the total number of predefined targets. For each environment, three targets are instantiated per quest stage using the schemas in Table~\ref{tab:targets}. The stage position $i$ indexes the stage inside a quest, while the concrete target object and required action depend on the corresponding quest definition.

\begin{table}[h!tp]
    \centering
    \caption{Predefined target schemas per environment and quest stage}
    \label{tab:targets}
    \scriptsize
    \setlength{\tabcolsep}{2pt}
    \renewcommand{\arraystretch}{1.08}
    \begin{tabularx}{\textwidth}{@{}p{1.85cm}p{1.55cm}>{\raggedright\arraybackslash}X>{\raggedright\arraybackslash}X@{}}
    \toprule
    \textbf{Environment} & \textbf{Target ID} & \textbf{Triggering condition (state or state-action)} & \textbf{Selection rationale} \\
    \midrule
    Hero Game & HG-S$i$-T1 & state: the player reaches the target object or prerequisite interaction state specified by stage $i$ (NPC, monster, item, or special terrain) & verifies access to the stage-specific interaction context \\
    Hero Game & HG-S$i$-T2 & state-action: the action required by stage $i$ (talk / attack / use / submit / terrain interaction) is executed on the corresponding target object & verifies the executability of the required stage operation \\
    Hero Game & HG-S$i$-T3 & state: stage $i$ is completed and the quest list advances to stage $i{+}1$ & verifies stage transition and quest progression \\
    Kitchen Game & KG-S$i$-T1 & state: the ingredient or intermediate object required by stage $i$ is acquired or placed at the relevant processing location & verifies access to the resource required for the stage \\
    Kitchen Game & KG-S$i$-T2 & state-action: the processing or manipulation action required by stage $i$ (pick up / place / cut / plate / submit) is executed on the correct object and location & covers stage-specific kitchen operations and state management \\
    Kitchen Game & KG-S$i$-T3 & state: the intermediate product or final dish required by stage $i$ is completed, and the quest advances or terminates & verifies intermediate progression or final dish submission \\
    Minecraft & MC-S$i$-T1 & state: the prerequisite resource, tool, location, or entity required by stage $i$ is obtained or reached & verifies collection, crafting prerequisites, and navigation \\
    Minecraft & MC-S$i$-T2 & state-action: the stage-required mine / craft / attack / build / enter action is executed on the corresponding target or at the required location & covers the operation that realizes the stage objective \\
    Minecraft & MC-S$i$-T3 & state: the stage-$i$ objective is reached (e.g., an item is crafted, a portal is entered, or the target mob is defeated) & verifies stage completion \\
    \bottomrule
    \end{tabularx}
\end{table}

\section{Reproducibility Settings}\label{app:repro}

Table~\ref{tab:llm_settings} reports the API settings of every LLM. None of the APIs used in this study expose deterministic model snapshots or seeds; the API model identifier returned with each response is recorded in the logs. Table~\ref{tab:seeds} summarizes the run-level seeds and sources of stochasticity for all methods.

\begin{table}[h!tp]
    \centering
    \caption{LLM API settings used by LLMEGQT}
    \label{tab:llm_settings}
    \scriptsize
    \setlength{\tabcolsep}{2pt}
    \renewcommand{\arraystretch}{1.1}
    \begin{tabularx}{\textwidth}{@{}p{2.4cm}p{1.4cm}>{\raggedright\arraybackslash}X>{\raggedright\arraybackslash}p{2.7cm}@{}}
    \toprule
    \shortstack{\textbf{Model}\\\textbf{(API model ID)}} & \textbf{Provider} & \shortstack{\textbf{Reasoning}\\\textbf{mode}} & \shortstack{\textbf{Generation}\\\textbf{settings}\\Temp.; top-$p$\\max. tokens; samples} \\
    \midrule
    gpt-5 (gpt-5) & OpenAI & built-in (default) & 0.7; 1.0; 2048; 1 \\
    gpt-4.1 (gpt-4.1) & OpenAI & none & 0.7; 1.0; 2048; 1 \\
    o4-mini (o4-mini) & OpenAI & reasoning effort ``medium'' & 1.0; 1.0; 2048; 1 \\
    qwen3-plus (qwen3-plus) & Alibaba & default & 0.7; 1.0; 2048; 1 \\
    deepseek-v3 (deepseek-v3) & DeepSeek & default & 0.7; 1.0; 2048; 1 \\
    \bottomrule
    \end{tabularx}
\end{table}

\begin{table}[h!tp]
    \centering
    \caption{Run-level seeds and sources of stochasticity}
    \label{tab:seeds}
    \footnotesize
    \setlength{\tabcolsep}{2pt}
    \renewcommand{\arraystretch}{1.08}
    \begin{tabularx}{\textwidth}{@{}p{2.4cm}>{\raggedright\arraybackslash}X@{}}
    \toprule
    \textbf{Method} & \textbf{Repeated-run setting} \\
    \midrule
    Random & ten runs per configuration using seeds $\{100, 101, \dots, 109\}$ \\
    GA & one seed per run from $\{100, 101, \dots, 109\}$; tournament selection \\
    CD-PPO & one seed per run from $\{100, 101, \dots, 109\}$; evaluation uses the greedy policy \\
    LLMEGQT & run-to-run variation arises solely from LLM sampling under the settings in Table~\ref{tab:llm_settings}; no separate PRNG seed or randomized exploration schedule is used \\
    \bottomrule
    \end{tabularx}
\end{table}

\end{document}